\documentclass[12pt,a4paper,oneside]{article}
\usepackage{amsfonts,amssymb,amsmath,amsthm,graphicx}

\title{{Casimir energy 
\\and dilute dielectric ball}}
\author{Valery N. Marachevsky \thanks{E-mail: root@VM1485.spb.edu}}
\date{{\normalsize Department of Theoretical Physics\\St.Petersburg State 
University \\198904 St.Petersburg , Russia}\\  }
 
\begin{document}
\newcommand{\om}{\bigl|\omega\bigr|\sqrt{\varepsilon}}
\newcommand{\omm}{\bigl|\omega\bigr|}

\maketitle

\begin{abstract}
\qquad General formalism of quantum field theory and addition
theorem for Bessel functions are applied 
to derive formula for Casimir-Polder
energy  of interaction  between a  polarizable particle  and  a dilute
dielectric ball.  The equivalence of dipole-dipole interaction
and Casimir energy for dilute homogeneous dielectrics is shown.
A novel method is used to derive
Casimir energy  of a dilute dielectric  ball 
without divergences in calculations.
Physically realistic model of a dilute ball
is discussed. Different approaches  to the
calculation of Casimir  energy of a dielectric  ball  are reviewed.
\end{abstract}

\begin{flushleft}
\end{flushleft}

\newpage

\section{Introduction}
The  study of  spherical geometry  in Casimir  effect meets  a  lot of
technical  problems,  the  most   difficult  one  is  the  problem  of
divergences, which appear in  many expressions.  The special interest
to the subject raised after the series of articles by Julian Schwinger
where he had suggested a connection between  Casimir  effect and
sonoluminescence   \cite{Schwinger,Sono}.   Different regularizations
were used to obtain finite results in spherical geometry,
however, results of the  calculations strongly depended
on the regularization scheme and regularization parameters
(a discussion of divergences and their regularization can be found
in \cite{Vassilevich}).
In the pesent paper we show that it is necessary that 
regularization parameters be present even in final expressions,
they have real physical meaning and we will clarify it. 

We study a dielectric nonmagnetic  ball of radius $a$ and permittivity
$\varepsilon$, surrounded  by a vacuum.  The ball is dilute,  i.e. all
final expressions  are obtained under the assumption  $\varepsilon - 1
\ll 1$. Dispersion is essential in the following, so 
the permittivity $\varepsilon$ depends on frequency.   

In section $2$ we follow  the formalism which  was developed by
E.Lifshitz  et.al.\cite{Lifshitz}  and K.Milton  et.al.\cite{Milton1}.
We start  from a short  overview of known  facts.  Then we  derive 
Casimir-Polder type  energy of interaction 
between  a dielectric  ball and  a  particle of
constant polarizability $\alpha$, which  is placed at the distance 
$r > a$ from the centre of the ball.
The addition theorem for Bessel functions
is used.  Another application of this technique can be found
in \cite{Mar3} where Casimir energy of polarizable particle
located in the neighbourhood of perfectly conducting wedge
was calculated.
In section $3$ the equivalence of dipole-dipole
interaction and Casimir energy in the lowest order
$(\varepsilon - 1)^2$ is proved for homogeneous dielectric media. 
This fact is used in section $4$ to calculate 
non-dispersive contribution to the energy, dispersive
contribution is briefly discussed. We argue that 
Casimir surface force is attractive in a 
realistic  model of dielectric ball  and includes 
volume and surface terms in the order $(\varepsilon - 1)^2$. 
 
Various regularizations are reviewed in the
Appendix.

We put $\hbar=c=1$. Heaviside-Lorentz units are used.

\newpage

\section{Polarizable particle and a ball} 
 
The change  in the  ground state  energy $E$ of  the system  under the
infinitesimal variation of $\varepsilon$ is
\begin{equation}
\delta  E  =\frac{i}{2}  \int  d^3  {\bf  x}  \int_{-\infty}^{+\infty}
\frac{d\omega}{2 \pi} \,  \delta\varepsilon ({\bf x},\omega) \,
D_{pp}(\varepsilon, {\bf r}, {\bf r}, \omega) \, . \label{f1}
\end{equation}
Here $D_{jk}(\varepsilon, {\bf r}, {\bf r}, \omega )$ 
is a Fourier component of electric field propagator trace, it 
satisfies equations
\begin{equation} 
\Bigl[ \varepsilon ({\bf r} , \omega ) \, \omega^2 \delta_{j m} -
{\rm  rot}_{j l}{\rm rot}_{l m} \Bigr] 
D_{m k}(\varepsilon, {\bf r},{\bf r}^\prime, \omega) = 
- \omega^2 \delta ({\bf r} - {\bf r}^\prime) \delta_{j k}
\end{equation}
The system
of equations  for this Green's  function was discussed  extensively in
\cite{Lifshitz,Milton1} (there is a discussion of these equations
in the next section). 
 The solution  of this  system  for spherical
geometry  with  standard boundary  conditions  classically imposed  at
$r=a$  can be  written  as in  \cite{Milton3} ($\delta$-functions  are
omitted  since  we are  interested  in the  limit  ${\bf  r} \to  {\bf
r}^\prime$):
\begin{multline}
D_{j k}(\varepsilon, {\bf r},{\bf r}^\prime, \omega)   =
-  \sum_{l=1}^{\infty}\sum_{m=-l}^{l}        (\omega^2
F_l(r,r^\prime)   X_{j l m}(\Omega)    X_{k l m}^*(\Omega^\prime)   +   \\
+\frac{1}{\varepsilon}{\rm  rot}_{{\bf  r}}{\rm rot}_{{\bf  r}^\prime}
G_l(r,r^\prime) X_{j l m}(\Omega) X_{k l m}^*(\Omega^\prime) ) .
\end{multline}
Here  we  have used  the  following  notations ($X_{ilm}(\Omega)$  are
vector  spherical harmonics;  $j_l (r),  h_l^{(1)} (r)$  are spherical
Bessel functions, $  \tilde e_l(r)=r h_l^{(1)}(r) , \tilde  s_l(r) = r
j_l (r) $ are Riccati-Bessel functions \cite{Jackson}):
\begin{equation}
X_{ilm}(\Omega)     =     \frac{1}{\sqrt{l(l+1)}}    ({\rm{\bf     L}}
Y_{lm}(\Omega))_i
\end{equation}
\begin{equation}
F_l, G_l = \left\{
\begin{array}{ll}
i k j_l  (k r_<) [h_l^{(1)} (k  r_>) - A_{F,G} j_l (k  r_>)], k=\om, &
r,r^\prime  < a,  \\  i \omm  h_l^{(1)}  (\omm r_>)  [j_l(\omm r_<)  -
B_{F,G} h_l^{(1)} (\omm r_<)], & r,r^\prime > a,
\end{array}
\right. \label{f4}
\end{equation}
\begin{gather}
A_F    =\frac{\tilde   e_l(\om    a)\tilde   e_l^\prime(\omm    a)   -
\sqrt{\varepsilon}    \tilde    e_l(\omm    a)\tilde    e_l^\prime(\om
a)}{\Delta_l}, \\ B_F  = \frac{\tilde s_l(\om a)\tilde s_l^\prime(\omm
a)  -  \sqrt{\varepsilon}   \tilde  s_l(\omm  a)\tilde  s_l^\prime(\om
a)}{\Delta_l}, \\ A_G =\frac{\sqrt{\varepsilon}\tilde e_l(\om a)\tilde
e_l^\prime(\omm   a)  -   \tilde   e_l(\omm  a)\tilde   e_l^\prime(\om
a)}{\tilde  \Delta_l}, \\ B_G  =\frac{\sqrt{\varepsilon}\tilde s_l(\om
a)\tilde s_l^\prime(\omm a)  - \tilde s_l(\omm a)\tilde s_l^\prime(\om
a)}{\tilde \Delta_l},
\end{gather}
\begin{eqnarray}
\Delta_l  &   =  &  \tilde  s_l(\om  a)\tilde   e_l^\prime(\omm  a)  -
\sqrt{\varepsilon} \tilde  s_l^\prime (\om a)\tilde e_l (\omm  a) , \\
\tilde  \Delta_l  & =  &  \sqrt{\varepsilon}  \tilde s_l(\om  a)\tilde
e_l^\prime(\omm a) - \tilde s_l^\prime (\om a)\tilde e_l (\omm a),
\end{eqnarray}
differentiation is taken over the whole argument.

When the  point particle of constant polarizability $\alpha$ is
inserted into
the point  ${\bf r}$, $|{\bf r}|>a$  from the centre of  the ball, the
energy  change is  given by  (\ref{f1}) with  $\delta\epsilon =  4 \pi
\alpha  \delta^3({\bf r}-  {\bf x})$.   However,  the contact
terms (empty space contribution)
 have to be subtracted from (\ref{f1}), i.e. when 
physical  quantities  are calculated in the 
region  $r, r^\prime  >  a$  the  term $i\omm h_l^{(1)}(\omm  r_>) j_l(\omm
r_<)$ has to be subtracted 
from (\ref{f4}) (for $r, r^\prime  < a$ in full analogy the term
$i  k   j_l(k  r_<)  h_l^{(1)}(k  r_>)$  should   be  subtracted  from
(\ref{f4})).  Doing  so, we  have to substitute  $\tilde F_l  , \tilde
G_l$ instead of $F_l, G_l$ in all expressions, where
\begin{equation}
\tilde F_l, \tilde G_l = \left\{
\begin{array}{ll}
 - i A_{F,G} k j_l (k r_<) j_l  (k r_>), k=\om, & r,r^\prime < a, \\ 
 - i B_{F,G}  \omm   h_l^{(1)}  (\omm  r_>)  h_l^{(1)}   (\omm  r_<),  &
 r,r^\prime > a.
\end{array}
\right.
\end{equation}
The Casimir-Polder energy of this configuration is
\begin{multline}
E_1 (r, a) = i \, \alpha \int_{-\infty}^{+\infty}d\omega 
D_{pp} (\varepsilon, {\bf r}, {\bf r}, \omega)   =   \frac{\alpha}{i}
\int_{-\infty}^{+\infty}   d\omega  \sum_{l=1}^{\infty}  \frac{2l+1}{4
\pi} \times  \\ \times \Bigl(\omega^2 \tilde  F_l(r,r^\prime) + l(l+1)
\frac{\tilde     G_l(r,r^\prime)}{r     r^\prime}     +     \frac{1}{r
r^\prime}\frac{\partial}{\partial           r}r\frac{\partial}{\partial
r^\prime} (r^\prime  \tilde G_l(r,r^\prime))\Bigr)\Bigr|_{r^\prime \to
r} \, .\label{f10}
\end{multline}
We perform a Euclidean rotation then: $\omega \to i\omega$ ,
\begin{equation}
\tilde s_l  (x) \to  s_l (x) =  \sqrt{\frac{\pi x}{2}}  I_{l+1/2} (x),
\tilde e_l (x) \to e_l (x) =\sqrt{\frac{2 x}{\pi}} K_{l+1/2} (x).
\end{equation}   
Let $x=\omega a$ . For $E_1(r, a)$ we obtain
\begin{multline}
E_1    (r,    a)=    \frac{2    \alpha}{a}    \int_{0}^{+\infty}    dx
\sum_{l=1}^{+\infty}  \frac{2 l+1}{4  \pi} \Bigl[  \frac{x}{a  r^2} \,
e_l^2(x  r/a) B_F  - \\  -\frac{l(l+1)  a}{r^4 x}  \, e_l^2(xr/a)  B_G
-\frac{x}{r^2 a}\, (e_l^\prime (xr/a))^2 B_G \Bigr].  \label{f9}
\end{multline} 
This expression  can be transformed to  a simple formula  in the limit
$\varepsilon  -   1\ll  1$.   The   functions  $B_F$  and   $B_G$  are
proportional to  $(\varepsilon -  1)$ in this  limit. To  proceed, the
following  addition  theorem  for  Bessel  functions  \cite{Abram}  is
useful:
\begin{gather}
u (p, k, x, \rho) \equiv  \sum_{l=0}^{+\infty} (2l+1) s_l (x p) e_l (x
k)  P_l  (\cos\theta) =  \frac{x  e^{- x\rho}  p  k}{\rho}  , \\  \rho
=\sqrt{p^2+k^2-2 p k \cos\theta}.
\end{gather}
To  our  knowledge this  formula  was  first  used in  Casimir  effect
calculations  in   \cite{Klich},where  it  was   applied  to  analytic
calculation of Casimir energy  of perfectly conducting spherical shell
and dilute dielectric ball satisfying $\varepsilon \mu = 1$.

In our case  it can be applied as follows.   The simple identity holds
($ k > p > 0 $ for definiteness):
\begin{multline}
\int dx \sum_{l=0}^{+\infty} (2l+1) f(x) s_l (x p) e_l (x k) s_l (x p)
e_l (x k)  = \\ =\frac{1}{2} \int_{k- p}^{k+p}  \frac{d\rho \, \rho}{p
k} \int dx f(x) u (p, k, x, \rho) u (p, k, x, \rho),
\label{f17}
\end{multline} 
where we have used
\begin{gather}
\int_{-1}^{1}   d(\cos\theta)  P_l   (\cos\theta)   P_m(\cos\theta)  =
\frac{2}{2l+1} \, \delta_{lm} , \\ \int_{-1}^{+1}d(\cos\theta)\cdots =
\int_{k-p}^{k+p} \frac{d\rho \, \rho}{pk}\cdots \,\, .  \label{ff1}
\end{gather}
Only the first order $\sim (\varepsilon - 1)$ is needed in $E_1$. We put
$k=r/a, \, p=1$ and use (\ref{f17}) and its obvious generalizations in
(\ref{f9}) to calculate $E_1$.  Finally we get
\begin{equation}
E_1 (r, a)= - \frac{23}{15} \, \alpha \, \frac{\varepsilon - 1}{4 \pi}
\,\frac{ a^3 (5 r^2 + a^2)}{r (r+a)^4 (r-a)^4}\, , r>a \,.
\label{g1}
\end{equation} 
Substitution  $\varepsilon - 1  = 4  \pi N_{mol}\alpha_{ball}$  in the
limit $r \gg a$ yields
\begin{equation}
E_1  (r, a)\Bigl|_{r\gg  a} =  N_{mol}  \Bigl(\frac{4\pi a^3}{3}\Bigr)
\frac{-23 \alpha \alpha_{ball}}{4  \pi r^7} = N_{mol} \Bigl(\frac{4\pi
a^3}{3}\Bigr) E_{Cas-Pol}\, .
\label{f20}
\end{equation}
Thus  in this limit  the famous  Casimir-Polder energy  of interaction
between  two  polarizable particles  $E_{Cas-Pol}$  \cite{Cas} can  be
obtained directly from (\ref{f20}).

Certainly, this is not simply a coincidence, and 
next section is devoted to the 
proof of equivalence of dipole-dipole interaction
and Casimir energy in the order $(\varepsilon - 1)^2$.

\section{Dipole interaction and Casimir energy} 
We define differential operator
\begin{equation}
L_{jm} ( \varepsilon, {\bf r}, i \omega ) =
\Bigl[ \varepsilon ({\bf r} , i |\omega| ) \, \omega^2 \delta_{jm}  +
{\rm  rot}_{jl}{\rm rot}_{lm} \Bigr]  \, .  \label{p21} 
\end{equation}
Equations for electric field propagator 
are well known due to works
\cite{Lifshitz,Milton1} :
\begin{equation}
L_{jm} ( \varepsilon, {\bf r}, i \omega )
D_{mk}(\varepsilon, {\bf r},{\bf r}^\prime, i \omega)
 = - \omega^2 \delta ({\bf r} - {\bf r}^\prime) 
\delta_{jk} \, . \label{p22}
\end{equation} 
Suppose that $\varepsilon$ has changed in the point ${\bf r}_2$.
The changed system satisfies equation   
\begin{multline}
L_{jm} ( \varepsilon + \delta \varepsilon (i |\omega|) 
\, \delta ({\bf r}_2 - {\bf r})  , {\bf r}, i \omega ) \,
{\stackrel{\thicksim}{D}}_{mk}(\varepsilon + 
\delta \varepsilon (i |\omega|) \, \delta ({\bf r}_2 -{\bf r})
, {\bf r},{\bf r}^\prime, i \omega)
= \\ = - \omega^2 \delta ({\bf r} - {\bf r}^\prime) 
\delta_{jk}   \label{p23}
\end{multline}
Let us define $\delta D_{jk} ({\bf r},{\bf r}^\prime, i \omega)$ as
a term proportional to $\delta \varepsilon (i |\omega|)$ in
the expansion
\begin{equation}
{ \stackrel{\thicksim}{D} }_{jk} =  D_{jk} +
\delta D_{jk} ({\bf r}, {\bf r}^\prime, i \omega )
 + O( (\delta \varepsilon (i |\omega|))^2 )  \label{p24}
\end{equation}
\begin{figure}
\centering \includegraphics[height=8cm]{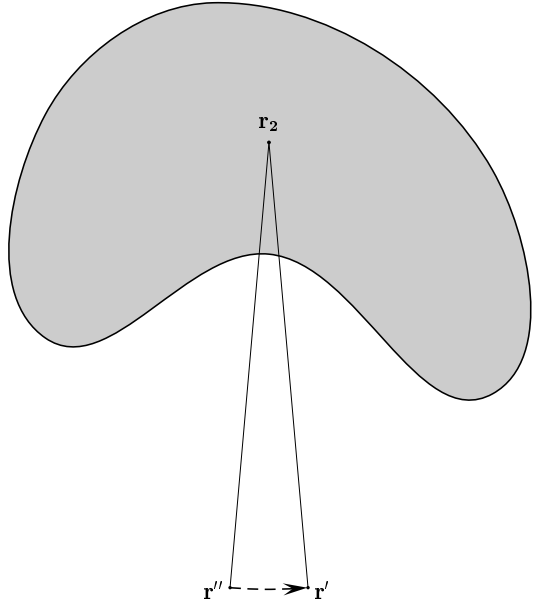}
\caption{dielectric area is shaded; point particle is located
  at point ${\bf r}^\prime$ }
\end{figure}
If we consider equation (\ref{p23}) and write series
in powers of constant $\delta \varepsilon (i |\omega|)$, 
in the first order we get the equality: 
\begin{equation}
\delta \varepsilon (i |\omega|) \, \delta ({\bf r}_2 - {\bf r}) \,
\omega^2 D_{jk} (\varepsilon, {\bf r},{\bf r}^\prime, i \omega) +
L_{jm} ( \varepsilon, {\bf r}, i \omega )
\delta D_{mk} ({\bf r}, {\bf r}^\prime, i \omega ) = 0 \, . \label{p25}
\end{equation}
We multiply equation (\ref{p25}) by $D_{pj} ({\bf r}^{\prime\prime},
{\bf r}, i \omega)$ on the left , integrate over ${\bf r}$ and
sum over $j$. Using properties $D_{pj} ({\bf r}^{\prime\prime},
{\bf r}, i \omega) = D_{jp} ( {\bf r}, {\bf r}^{\prime\prime}, i \omega)$
, $L_{jm}^{\dagger} = L_{mj}$ and equation (\ref{p22}) it is 
straightforward to obtain  
\begin{equation}
\delta D_{pk} ({\bf r}^{\prime\prime}, {\bf r}^\prime , i \omega) =
D_{pj} (\varepsilon, {\bf r}^{\prime\prime}, {\bf r}_2 , i \omega) 
\delta\varepsilon (i |\omega|) 
D_{jk} (\varepsilon, {\bf r}_2 , {\bf r}^\prime , i \omega).  \label{p26}
\end{equation}

Now suppose that 
$D_{mk}$ satisfies (\ref{p22}) with $\varepsilon=1$ in all space.
Then we insert dielectric of arbitrary shape with constant
dielectric permittivity $\varepsilon (i |\omega|)$ so that
$\delta \varepsilon (i |\omega|) = \varepsilon (i |\omega| ) - 1 $.
For this case $\delta$-function in (\ref{p25})
has to be substituted by function equal to $0$ outside the 
dielectric and equal to $1$ inside the dielectric.
Denoting
$D_{pk}^{(0)} ({\bf r}^{\prime\prime}, {\bf r}^\prime
 , i \omega) \equiv
 D_{pk} (\varepsilon=1, {\bf r}^{\prime\prime}, {\bf r}^\prime
 , i \omega)$ , 
we get the formula
\begin{align}
\begin{split}
&D_{pk} (\varepsilon, {\bf r}^{\prime\prime}, {\bf r}^\prime , i \omega)
- D_{pk}^{(0)} ({\bf r}^{\prime\prime}, {\bf r}^\prime , i \omega)
= \\& =
\int d^3 {\bf r}_2 \,
D_{pj}^{(0)} ({\bf r}^{\prime\prime}, {\bf r}_2 , i \omega) 
(\varepsilon (i |\omega|) - 1) 
D_{jk}^{(0)} ({\bf r}_2 , {\bf r}^\prime , i \omega) 
+ \, O( (\varepsilon(i |\omega|)-1)^2 )  \label{p27} 
\end{split}
\end{align}
Figure 1 illustrates formula (\ref{p27}).
When ${\bf r}^{\prime\prime} \Rightarrow {\bf r}^{\prime} $
the formula for Casimir energy of interaction between
a polarizable particle located at point ${\bf r}^\prime$ 
and dielectric can be written as 
\begin{equation}
\begin{split}
E &= - \int_{-\infty}^{+\infty} d\omega \, \alpha_1
 ({\bf r}^\prime, i |\omega|) 
 ( D_{pp} (\varepsilon, {\bf r}^{\prime\prime}, 
 {\bf r}^\prime, i \omega) -
  D_{pp}^{(0)} ({\bf r}^{\prime\prime}, {\bf r}^\prime, i\omega)  )
\Bigr|_{ {\bf r}^{\prime\prime} \to {\bf r}^\prime } = \\
 &= \int d^3 {\bf r}_2 \,\rho({\bf r}_2) 
U ({\bf r}^\prime , {\bf r}_2) 
+ \, O(\alpha^3) ,  \label{p28}
\end{split}
\end{equation}
where
\begin{equation}
U({\bf r}^\prime , {\bf r}_2) =
- \int_{-\infty}^{+\infty} d\omega \, 4\pi \alpha_1 ({\bf r}^\prime, i |\omega|)
\alpha_2 ({\bf r}_2, i |\omega|) \,
(D_{pj}^{(0)} ({\bf r}^\prime , {\bf r}_2 , i \omega))^2 
 . \label{p29} 
\end{equation} 

Equation for potential (\ref{p29}) can also be obtained from
dipole-dipole interaction  of 
two neutral atoms \cite{Dzialoshinskii}. We refer the
reader to this book. If we denote $r=|{\bf r}_2-{\bf r}^\prime|$, then
(see \cite{Dzialoshinskii})
\begin{equation}
U(r) = - \frac{1}{\pi r^2} \int_{0}^{+ \infty} \, 
\omega^4 \alpha_1(i \omega) \, \alpha_2 (i \omega)\,  e^{-2\, \omega r }
\Bigl[1+ \frac{2}{\omega r} + \frac{5}{(\omega r)^2 }
+ \frac{6}{(\omega r)^3} + \frac{3}{(\omega r)^4} \Bigr] d\omega  \label{p30}
\end{equation}  
If we define atomic absorption frequency 
as $\omega_{0}$,  then 
$\lambda \sim 1/\omega_{0}\,$ is a characteristic 
absorption wavelength in the
spectrum of interacting atoms.
For distances $r \gg \lambda$ the main contribution to (\ref{p30})
gives the  frequency region $\omega \lesssim 1/r \ll 1/\lambda$. 
In this frequency region according to general properties
of dielectric permittivity on imaginary axis it is poissible 
to use $\alpha_1 (0)$ and $\alpha_2 (0)$ instead of 
$\alpha_1 (i \omega)$ and $\alpha_2 (i \omega)$,
so (\ref{p30}) for $r \gg \lambda$ yields Casimir-Polder
potential 
\begin{equation}
U(r)\Bigl|_{r\gg \lambda} = - \frac{23}{4\pi} \frac{\alpha_1 (0)\,
  \alpha_2 (0)}{r^7} . \label{p31}
\end{equation}  
For $\lambda_{at.} \ll r \ll \lambda$  ($\lambda_{at.}$ is atomic size)
frequencies $\omega \sim 1/\lambda$ give
essential contribution to (\ref{p30}). Then the last term in
(\ref{p30}) gives main contribution and 
\begin{equation}
U(r)\Bigl|_{r \ll \lambda} = -\frac{3}{\pi r^6}
\int_{0}^{+\infty}
d\omega \, \alpha_1(i \omega) \, \alpha_2(i \omega) \, .\label{p32} 
\end{equation}
Formulas (\ref{p31}) and (\ref{p32}) are non-dispersive and 
dispersive contributions to the energy of interaction
of two neutral atoms.

\section{Energy of a dilute ball}
The change in the ground state energy of a dilute dielectric ball 
is generally equal 
\begin{equation}
dE = \frac{dE}{dV} \Bigl|_{\varepsilon=const} dV +
 \frac{dE}{d\varepsilon}\Bigl|_{V=const} d\varepsilon \, ,  
\label{p1} 
\end{equation}
where $V$ is the volume of the ball.
Usually only the first term had been taken into account
in the calculations of Casimir force and, as a consequence,
the number of particles inside the ball wasn't conserved and 
the repulsive force resulted in most calculations.
However, from physical considerations
all atoms constituting the ball interact
at large distances by attractive Casimir-Polder potential,
so non-dispersive component of the force should be attractive.  
 
Conservation of the number of atoms inside the ball 
is equivalent to the conditions 
\begin{equation}
  (\varepsilon - 1) V = const, \qquad \quad 
 d\varepsilon =  - \frac{(\varepsilon - 1)\, dV}{V}    \label{p2}
\end{equation}   

We start from the first term contribution to (\ref{p1})
which we denote $dE_1$. 
\begin{figure}
\centering \includegraphics[height=6cm]{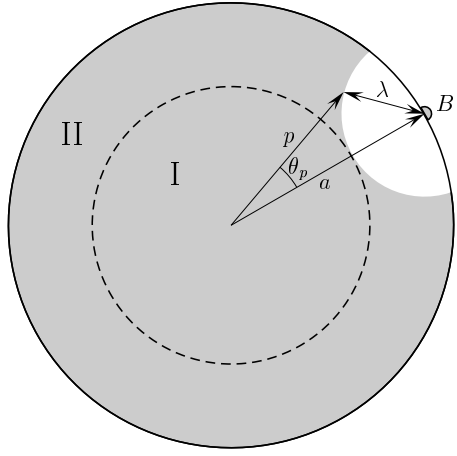}
\caption{contribution to $dE_{1 \, non-disp.}$}
\end{figure}
Figure 2 clarifies designations. 
When particles are inserted into the 
(infinitely small) shaded area near point $B$,
they interact via Casimir-Polder (retarded) potential
with particles inside a ball of radius $a$ separated 
by distances $r \gtrsim \lambda$
from the point $B$, i.e. they interact via Casimir-Polder 
potential with particles inside 
the shaded region of the ball (regions $I+II$). 
Particles separated from the point $B$
by distances $\lambda_{at.} <  r_{interatomic} \ll  r  \lesssim \lambda$
interact with particles near point $B$ via van der Waals 
(non-retarded) potential. 
Obviously the situation is simplified for clarity here.  
Generally one has to take the concrete
model of dispersion for dielectric permittivity,
integrate interaction of particles 
near point $B$ via potential (\ref{p29}) with particles  
separated by distances greater than interatomic 
distances  from point $B$
and calculate the energy change according to formula (\ref{p1}).          
As we are not intending to study specific 
model of dispersion here but rather want to get a   
suitable approximation for low-frequency contribution 
to the energy,  
we approximate area of integration by its division
on the retarded region ($r \gtrsim \lambda$) 
and non-retarded region ($r_{interatomic} \lesssim  r  \lesssim \lambda$).
According to discussion in the previous section, for
distances $r \gtrsim \lambda$ the frequency dispersion can be neglected,  
so this region in our approximation gives non-dispersive contribution
to the energy , the region $r_{interatomic} \lesssim  r  \lesssim \lambda$
gives main dispersive contribution to the energy.

Area of integration for non-dispersive case is shaded on fig.2,
so in our approximation 
the non-dispersive contribution to $dE_1$ can be written as 
\begin{multline}
dE_{1 \, non-disp.} =  4\pi a^2 da \, \int\limits_{I+II}  d^3{\bf r}
\, \frac{-23 \, \alpha ({\bf r}_B,\omega =0) \, 
\rho ({\bf r}_B) \, \alpha ({\bf r},\omega =0) \, \rho ({\bf r})  } 
{4\pi |{\bf r}_B - {\bf r}|^7} = \\ 
= 4 \pi a^2 da \, \frac{(\varepsilon(\omega =0) - 1)^2}{(4 \pi)^2}
\int\limits_{I+II}  d^3{\bf r}
\, \frac{ -23 }{4\pi |{\bf r}_B - {\bf r}|^7}    ,    \label{p3}
\end{multline} 
where $\rho ({\bf r}_B) $ and $\rho (\bf r)$ are densities of particles
in a unit volume. The integral over the region $I$ can be simply
obtained from (\ref{g1}) and arguments of the previous section.
We just substitute $a \Rightarrow a-\lambda , \, r \Rightarrow a$ 
in formula (\ref{g1}) according to fig.2 and obtain:
\begin{equation}
\int\limits_{I}  d^3{\bf r}
\, \frac{ -23 }{4\pi |{\bf r}_B - {\bf r}|^7} =
-\frac{23}{15} \frac{(a-\lambda)^3 (5 a^2 + (a-\lambda)^2)}
{a (2a - \lambda)^4 \lambda^4}      \label{p4}
\end{equation}    
The integral (\ref{p3}) over the region $II$ can be calculated 
analytically using (\ref{ff1}):
\begin{multline}
\int\limits_{II}  d^3{\bf r}
\, \frac{ 1 }{|{\bf r}_B - {\bf r}|^7} =
\int\limits_{a-\lambda}^{a} dp \int\limits_{\theta_p}^{\pi} d\theta \,
\frac{2\pi p^2 \sin\theta}{(a^2+p^2-2ap \cos\theta)^{\frac{7}{2}}} =
\\=\int\limits_{a-\lambda}^{a} dp \, \frac{2\pi p}{a} 
\int\limits_{\lambda}^{a+p} d\rho \, \frac{1}{\rho^6} 
=\frac{1}{480}\frac{\pi (5\lambda^4 + 192 a^4 -96 a^3\lambda)}
{a^4 \lambda^4}  -\frac{1}{30} \frac{\pi(5a - 4\lambda)}
{a (2a - \lambda)^4}   \label{p5}
\end{multline} 
From (\ref{p3}), (\ref{p4}) and (\ref{p5}) after integration over $a$
(keeping $\varepsilon = \mbox{const}$)
we obtain expression for $E_{1 \, non-disp.}$:
\begin{equation}
E_{1 \, non-disp.} =
(\varepsilon - 1)^2 \Bigl(-\frac{23}{128}\frac{V}{\pi^2 \lambda^4}
+ \frac{23}{384}\frac{S}{\pi^2 \lambda^3} + 
\frac{23}{1536} \frac{1}{\pi a} \Bigr)
\,\, + \, \mbox{const}   \label{p6}
\end{equation}  
Here $V$ and $S$  are volume and surface area of the ball respectively.
Expression (\ref{p6}) looks similar to formula (\ref{y3}) 
obtained by G.Barton \cite{Barton} in the framework
of quantum mechanical perturbation theory. There is still a 
little difference between these formulas.   
At first sight it seems that in our approach it is impossible to find the 
term proportional to $1/\lambda$ and this term
has no influence on the force.  However, this is not so.
The term proportional to $1/\lambda$ is determined uniquely 
in our approach when taking into account the second term in
(\ref{p1}) and equation (\ref{p2}). 

\begin{figure}
\begin{center}
\includegraphics[height=4cm]{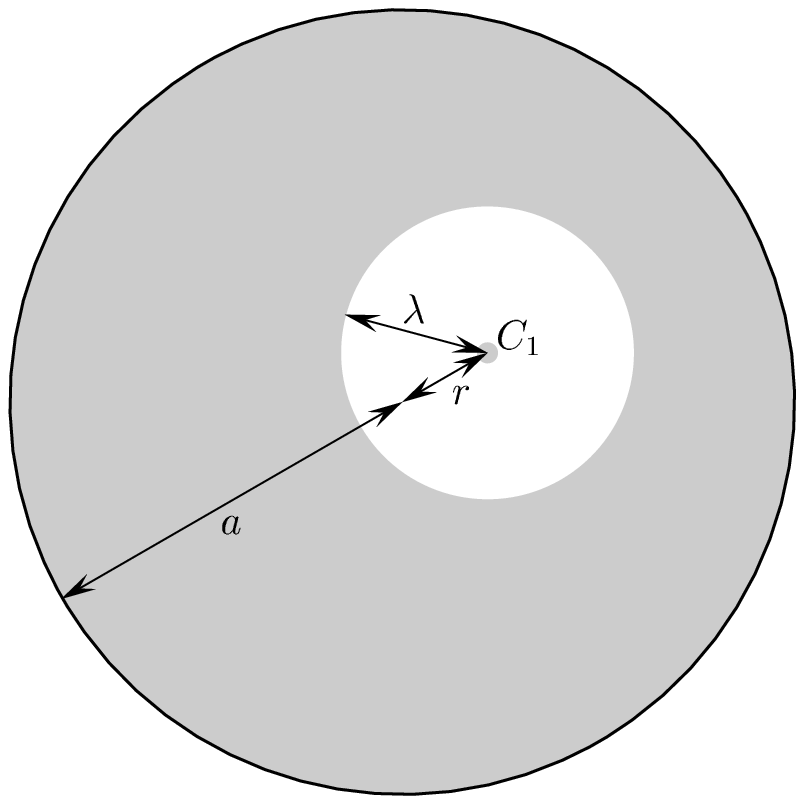} \hfill 
\includegraphics[height=4cm]{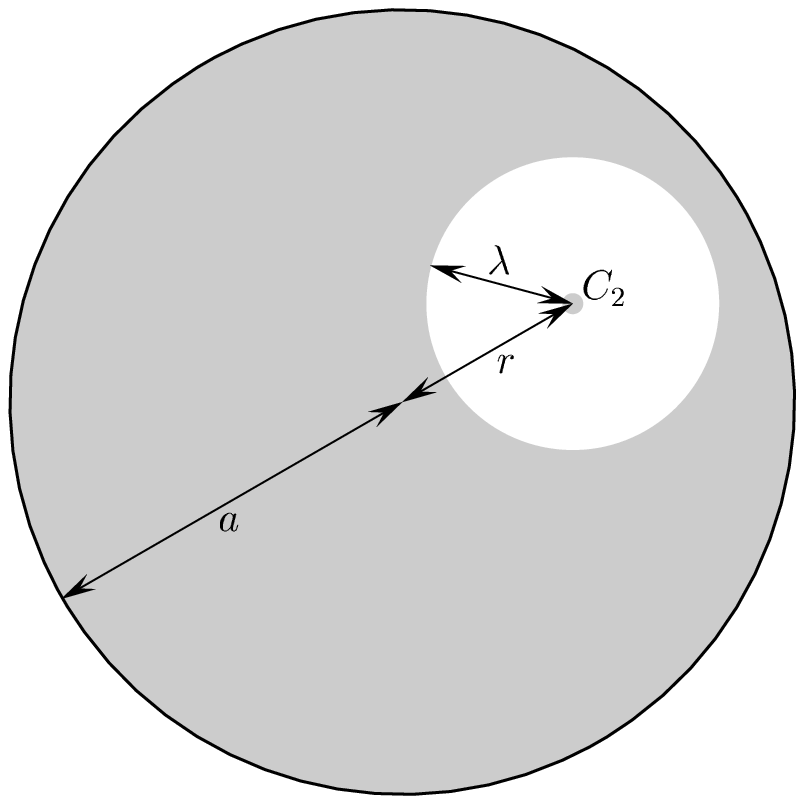} \hfill 
\includegraphics[height=4cm]{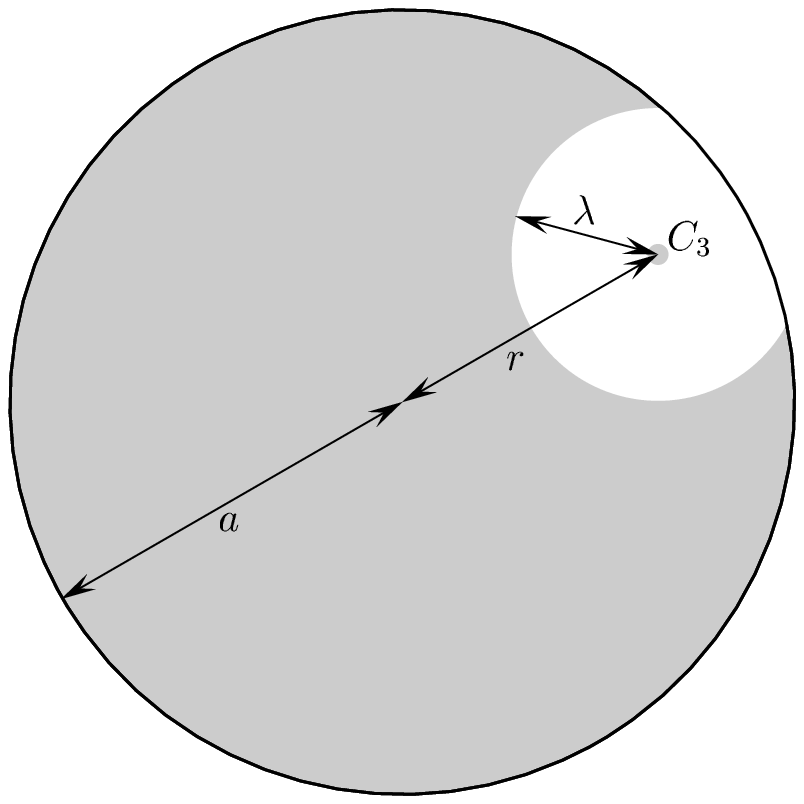} \hfill 

\vspace{0.5cm}
Figure 3: contributions to $dE_{2 \, non-disp.}$
\end{center}
\end{figure}
Now we turn to the second term in (\ref{p1}) and calculate 
its non-dispersive contribution to the energy change 
denoted by $dE_{2 \, non-disp.}$.   
Figure 3 illustrates different areas of integration over variable $r$ 
and analogous to fig.2 in a sence -- it is necessary 
to integrate Casimir-Polder potential between 
(infinitesimally small area around) point $C_1$ ($C_2$, $C_3$) and
shaded area of the ball separated from point $C_1$ 
($C_2$, $C_3$) by distances 
$r \gtrsim \lambda$ and then integrate over points $C_1$, $C_2$, $C_3$ 
inside the ball.  This term can be written as follows:
\begin{equation}
dE_{2 \, non-disp.} \Bigl|_{V=const} = \frac{\varepsilon - 1}{4\pi}
\frac{\delta \varepsilon}{4\pi}
\iint\limits_{|{\bf r}_C - {\bf r}| \ge \lambda}
dV ({\bf r}_C) dV ({\bf r})  \,
\frac{-23}{4\pi |{\bf r}_C - {\bf r}|^7} , \label{p7}
\end{equation} 
where  ${\bf r}_C$  and ${\bf r}$ are integrated over 
all points inside the ball of radius $a$ satisfying 
the condition $|{\bf r}_C - {\bf r}| \ge \lambda$. 

To facilitate understanding of calculations in what follows 
we present major steps explicitly. Let us denote 
\begin{equation}
g(r,p,\rho) = 4\pi r^2 \cdot 2\pi p^2 \frac{1}{rp \rho^6} \label{p8}
\end{equation} 
and write $dE_{2 \, non-disp.}$ 
(after comparison with (\ref{p5}) and fig.3 \, 
everything should be clear):  
\begin{multline}
dE_{2 \, non-disp.} = -\frac{23}{4\pi} \frac{\varepsilon-1}{4\pi}
\frac{\delta\varepsilon}{4\pi} \times \\
\times \Biggl(  \biggl(\int\limits_{0}^{\lambda} dr 
\int\limits_{\lambda - r}^{\lambda + r} dp \int\limits_{\lambda}^{r+p} d\rho
\,  + \int\limits_{0}^{\lambda} dr 
\int\limits_{r+\lambda}^{a} dp \int\limits_{p-r}^{r+p} d\rho \biggr)
\, g(r,p,\rho) \qquad \mbox{see fig.3.1} \quad +  \\
 + \biggl(\int\limits_{\lambda}^{a - \lambda} dr 
\int\limits_{r+\lambda}^{a} dp \int\limits_{p-r}^{p+r} d\rho
\, + \int\limits_{\lambda}^{a-\lambda} dr 
\int\limits_{r-\lambda}^{r+\lambda} dp \int\limits_{\lambda}^{r+p} d\rho
\, + \\  +\int\limits_{\lambda}^{a-\lambda} dr 
\int\limits_{0}^{r-\lambda} dp \int\limits_{r-p}^{r+p} d\rho \biggr)
\, g(r,p,\rho)  \qquad
\mbox{see fig.3.2} \quad + \\
+ \biggl( \int\limits_{a-\lambda}^{a} dr 
\int\limits_{r-\lambda}^{a} dp \int\limits_{\lambda}^{r+p} d\rho
+ \int\limits_{a-\lambda}^{a} dr 
\int\limits_{0}^{r- \lambda} dp \int\limits_{r-p}^{r+p} d\rho \biggr)
g(r,p,\rho)  \qquad \mbox{see fig.3.3}  \Biggr) = \\
= 2 \delta\varepsilon (\varepsilon - 1) 
\Bigl(  -\frac{23}{128}\frac{V}{\pi^2 \lambda^4}
+ \frac{23}{384}\frac{S}{\pi^2 \lambda^3}
 - \frac{23}{384}\frac{1}{\pi\lambda} + 
\frac{23}{1536} \frac{1}{\pi a}  \Bigr).   \label{p9}
\end{multline}
The expression for non-dispersive force on a unit surface 
can be obtained  using (\ref{p1}) and (\ref{p2}):
\begin{multline}
F_{non-disp.} = - \frac{1}{4\pi a^2} \frac{dE}{da} = 
 - \frac{1}{4\pi a^2} \frac{dE_{1 \, non-disp.} 
+ dE_{2 \, non-disp.}}{da} = \\ 
= - \frac{23}{128 \pi^2 \lambda^4} \Bigl( 1 - \frac{4}{3} \frac{\lambda}{a}             
 + \frac{1}{2} \frac{\lambda^3}{a^3}
 - \frac{7}{48} \frac{\lambda^4}{a^4} \Bigr).   \label{p10}        
\end{multline}
This force is attractive for $\lambda \le a$ , 
which reflects attractive nature
of Casimir-Polder potential.

To calculate energy, one has to take into account equation (\ref{p2})
and integrate force. As a result,
\begin{equation}
E_{non-disp.} = (\varepsilon - 1)^2
\Bigl(  -\frac{23}{128}\frac{V}{\pi^2 \lambda^4}
+ \frac{23}{384}\frac{S}{\pi^2 \lambda^3}
 - \frac{23}{384}\frac{1}{\pi\lambda} + 
\frac{23}{1536} \frac{1}{\pi a}  \Bigr). \label{p11}
\end{equation} 
The term proportional to $1/\lambda$ is determined uniquely now.
Moreover, comparison of (\ref{p6}) and (\ref{p11}) shows internal
self-consistence of our approach.

\section{Conclusions}
The  formulas  for   Casimir-Polder  energy  of 
interaction between a
polarizable particle and a dilute dielectric ball (\ref{g1}),(\ref{f20})
and  non-dispersive contribution to 
Casimir surface  force on  a dilute  dielectric  ball (\ref{p10})
are derived analytically.  
Our approach directly  shows correspondence of
this force and Casimir-Polder potential.
In section $3$ this topic has been investigated from
general viewpoint for arbitrary homogeneous dielectric media.
The proof of equivalence in summation via Casimir-Polder - van der Waals
type potentials and Casimir energy in the
lowest $(\varepsilon - 1)^2$ order
is the subject of section $3$.    
This equivalence  made it possible to calculate 
non-dispersive contribution to Casimir energy 
of a dilute dielectric ball  by direct summation 
of Casimir-Polder potential.
When the number of atoms constituting dilute ball is assumed constant
during the ball expansion ,
the Casimir surface force is attractive and includes volume
and surface terms in the order $(\varepsilon - 1)^2$.

\section{Appendix}
For the benefit of the reader it is  helpful
to give here an overview of  different approaches
which have  been used  to derive the  finite
when $\lambda \to 0$ term  $E_f = 23/(1536 \pi a)$
in the energy expression (\ref{p11}) . 

First attempts to calculate Casimir energy of a
dielectric ball gave a lot of different answers, as it is described in
\cite{Barton}.   The  mathematical reason  for  some of 
these differences  was found in  the work  \cite{Mar1}, 
where by  use of Debye  expansion for
Bessel functions  and $\zeta$-function  the correct limits  on 
$E_f$ were established.
Later  in the article  \cite{Mar2}  the term $E_f$
was  calculated numerically with high accuracy,
which made it possible to  establish equivalence of Casimir effect and
retarded van der Waals (or Casimir-Polder) energy 
for the case of non-dispersive dielectric
ball.

Mutual retarded van der Waals energy for molecules inside the compact
sphere (ball) was calculated by  K.Milton and Y.Ng \cite{Milton2} and its
finite part after regularization in terms of gamma functions was first
given by $E_f$ in the work \cite{Milton2}.

In the article  \cite{Mar2} Casimir energy of  a dilute dielectric
ball  was studied via  the formula  which has  the following  form for
nondispersive case:
\begin{equation}
E_C=-\frac{(\varepsilon-1)^2}{8\pi                  a}\sum_{l=1}^\infty
(2l+1)\int_{0}^\infty dx\, x\frac{d}{dx}F_l(x), \label{y1}
\end{equation}
where
\begin{equation}
F_l(x)=        -\frac{1}{4}\left(\frac{d}{dx}(e_ls_l)\right)^2       -
x^2\Bigl(s^{\prime\,   2}-  s  s^{\prime\prime}\Bigr)\Bigl(e^{\prime\,
2}-e e^{\prime\prime}\Bigr) ,
\label{y4}
\end{equation}
and Riccati-Bessel  functions are assumed  to depend on  argument $x$.

The  formula (\ref{y1}) can  be studied  using formula  (\ref{f17}) as
well. We put $ k=1, p=1-m$, where we are interested in the limit $m\to
0$. For $E_C$ we find
\begin{equation}
E_C=\lim_{m\to         0}\frac{(\varepsilon         -        1)^2}{\pi
a}\left(\frac{23}{1536} + O \Bigl(\frac{1}{m}\Bigr) \right) \, .
\end{equation}

Recently  the  formula (\ref{y1})  has  been  studied in the mode
summation method \cite{mode}, there divergent terms have been analysed and
the finite result $E_f$ has also been derived by making use of addition
theorem for Bessel functions.

Another approach  based on quantum mechanical  perturbation theory was
suggested by G.Barton in  the work \cite{Barton}.
The Casimir  energy was obtained there in the form:
\begin{multline}
E=    -(\varepsilon    -    1)\frac{3}{2\pi^2}\frac{V}{\lambda^4}    +
\\+(\varepsilon                        -                        1)^2(-
\frac{3}{128\pi^2}\frac{V}{\lambda^4}+\frac{7}{360\pi^3}\frac{S}{\lambda^3}-
\frac{1}{20\pi^2}\frac{1}{\lambda}+  \frac{23}{1536\pi}\frac{1}{a}) \,
.\label{y3}
\end{multline}
Here $V$  is the volume  and $S$ the  surface area, $1/\lambda$  is an
exponential cutoff on wavenumbers.  In this approach the contact terms
haven't  been  subtracted,  this  is  why  the  term  proportional  to
$\varepsilon  - 1$ is  present in  (\ref{y3}).  We have already 
discussed this formula in section $4$, its structure is similar
to (\ref{p11}). 
The  cutoff independent term is essentially the same in all approaches.

The approach  based on quantum statistical mechanics  was developed in
\cite{Brevik}. This work led to similar results.

The  theory  of  QED  in   a  dielectric  background  was  studied  in
\cite{Vassilevich}    by    path    integral   and    $\zeta$-function
methods. $\zeta$-function can be written as in \cite{Vassilevich}:
\begin{equation}
\zeta(s)      =     \frac{\sin      \pi     s}{\pi}\sum_{l=1}^{\infty}
(2l+1)\int_{0}^{\infty}             d\omega             \,\omega^{-2s}
\frac{\partial}{\partial \omega} \ln ( \Delta_l \tilde \Delta_l ).
\end{equation} 
For a dilute ball it is  possible to expand the logarithm in powers of
$\varepsilon  - 1$,  here we  consider only  the order  $(\varepsilon -
1)^2$ of this expansion:
\begin{equation}
\zeta(s)    =   \frac{(\varepsilon    -   1)^2}{4}    \frac{\sin   \pi
s}{\pi}\sum_{l=1}^{\infty}       (2l+1)\int_{0}^{\infty}       d\omega
\,\omega^{-2s}   \frac{\partial}{\partial    \omega}   F_l(\omega   a)
,\label{y5}
\end{equation} 
where $F_l(x)$ is defined in (\ref{y4}).  
Detailed derivation of the finite part of $\zeta$ - function
deserves a single article , here for completeness
we only present the result for its finite part $\zeta_f$ :
\begin{equation}
\zeta_{f}(s) =  -(\varepsilon - 1)^2 \,\frac{\sin(\pi  s)}{\pi s} a^{2s}\,
2^{4s-7}\, \frac{(s^2-3s+4)\Gamma(-2s+2)}{(s-1)} ,
\end{equation}
with  no   poles  in  this   expression  for  $s<1$.  
From this expression we find
\begin{equation}
E_f = \frac{\zeta_f  (-1/2)}{2} = \frac{23}{1536} \frac{(\varepsilon
- 1)^2}{\pi a} \, ,
\end{equation}
we only note that even $\zeta$-function method meets
severe problems with analytic continuation and 
regularization in the case of a
dilute dielectric ball. 

Most divergences and problems met so far 
in different approaches are related to short distance behaviour. 
In this article it is shown explicitly that such problems arise
when parameter $\lambda \to 0$.
In dielectric media in the real world $\lambda \ne 0$,
this is why the main problem which still remains in other
approaches is not to remove regularization, but to
identify parameters of regularization in final expressions 
with physical quantities.

\end{document}